\documentclass[sigconf, nonacm]{acmart}
\usepackage{url}
\usepackage{makecell}
\usepackage{booktabs}
\usepackage{dcolumn}
\usepackage[T1]{fontenc}
\usepackage{hyperref}

\usepackage{bm}
\usepackage{multirow}

\AtBeginDocument{%
  }

\setcopyright{none}

\acmConference[Conference acronym 'XX]{Make sure to enter the correct
  conference title from your rights confirmation emai}{June 03--05,
  2018}{Woodstock, NY}

\newcommand{\AAMnotice}{%
\noindent\textbf{Author Accepted Manuscript (AAM).}\\
\textcopyright~2025 The Authors. Licensed under CC BY 4.0.\\
This is the author's version of the work. The definitive Version of Record
will be published in \emph{Proceedings of the 34th ACM International Conference on Information and Knowledge Management (CIKM~'25)}.
\acmISBN{978-1-4503-XXXX-X/18/06}
}

\begin{document}

\title{Effect of Model Merging in Domain-Specific Ad-hoc Retrieval}

\author{Taiga Sasaki}
\affiliation{%
  \institution{University of Hyogo}
  \city{Kobe}
  \country{Japan}
}
\email{ad24l026@guh.u-hyogo.ac.jp}

\author{Takehiro Yamamoto}
\affiliation{%
  \institution{University of Hyogo}
  \city{Kobe}
  \country{Japan}
}
\email{t.yamamoto@sis.u-hyogo.ac.jp}

\author{Hiroaki Ohshima}
\affiliation{%
  \institution{University of Hyogo}
  \city{Kobe}
  \country{Japan}
}
\email{ohshima@ai.u-hyogo.ac.jp}

\author{Sumio Fujita}
\affiliation{%
  \institution{LY Corporation}
  \city{Tokyo}
  \country{Japan}
}
\email{sufujita@lycorp.co.jp}

\renewcommand{\shortauthors}{Taiga Sasaki, Takehiro Yamamoto, Hiroaki Ohshima, and Sumio Fujita}

\begin{abstract}
In this study, we evaluate the effect of model merging in ad-hoc retrieval tasks. 
Model merging is a technique that combines the diverse characteristics of multiple models. 
We hypothesized that applying model merging to domain-specific ad-hoc retrieval tasks could improve retrieval effectiveness. 
To verify this hypothesis, we merged the weights of a source retrieval model and a domain-specific (non-retrieval) model using a linear interpolation approach.
A key advantage of our approach is that it requires no additional fine-tuning of the models.
We conducted two experiments each in the medical and Japanese domains. 
The first compared the merged model with the source retrieval model, and the second compared it with a LoRA fine-tuned model under both full and limited data settings for model construction.
The experimental results indicate that model merging has the potential to produce more effective domain-specific retrieval models than the source retrieval model, and may serve as a practical alternative to LoRA fine-tuning, particularly when only a limited amount of data is available. 
\end{abstract}


\maketitle
\AAMnotice

\section{INTRODUCTION}
Recent advancements in large language models (LLMs) have significantly improved the effectiveness of ad-hoc retrieval tasks~\cite{craswell2021trec}. 
In particular, retrievers based on LLMs with billions of parameters, such as those utilizing foundation models like Llama2 and Mistral, have demonstrated strong effectiveness~\cite{ma2024fine, wang-etal-2024-improving-text}.
 
However, retrievers fine-tuned on general-domain documents have limitations in improving effectiveness in specific domains~\cite{thakur2021beir}. This is because retrieval of domain-specific documents often requires understanding of specialized knowledge and unique expressions. 
Therefore, while constructing domain-specific retrievers is necessary, this process often incurs significant costs due to the need for additional domain-specific relevance judgment data or the computational cost of GPU resources for fine-tuning.

On the other hand, model merging, which combines multiple models to construct new foundation models, has recently gained attention. Model merging enables the creation of models that combine the capabilities of the source models. For instance, merging a model that has strong mathematical reasoning abilities with a model capable of code generation can result in a model proficient in both mathematical reasoning and code generation~\cite{yu2024language}.

So far, models constructed through model merging have primarily been utilized in text- and image-generation tasks. 
However, the application of model merging to ad-hoc retrieval tasks has not been explored. Therefore, this study aims to investigate whether model merging can be effective for ad-hoc retrieval tasks. 
Specifically, in this paper, we investigate whether model merging enables the construction of a domain-specific retriever.
For the merging process, we prepared two types of models: one fine-tuned for retrieval tasks and another that lacks retrieval capabilities but is further pre-trained for a specific domain. 
We then merged the weights of these two models by linear interpolation to construct a merged model, which is expected to perform better in the specific domain.

A key advantage of our approach is that it requires no additional fine-tuning of the models. 
One common approach for constructing domain-specific retrieval models is to prepare relevance judgment data tailored to the domain and fine-tune the model accordingly~\cite{xu2024bmretriever}. 
In contrast, our method builds a domain-specific retriever by merging an existing retrieval model with a model that has been further pre-trained on domain-specific data through a linear interpolation of their weights. 
This strategy can reduce the costs associated with fine-tuning.

The research questions of this study are as follows:
\textbf{RQ1:} Can model merging construct a domain-specific retrieval model that outperforms the source retrieval model?
\textbf{RQ2:} Is model merging a viable alternative to fine-tuning?
To answer these questions, we conducted two experiments\footnote{The code is available at \texttt{https://github.com/tyamamot/modelmerge-for-ir}.}, each conducted in the medical and Japanese domains. 
The first experiment compared the merged model with the source retrieval model.
The second experiment compared it with a LoRA fine-tuned model under both full and limited data settings for model construction.
The experimental results revealed that model merging has the potential to construct more effective domain-specific retrieval models than the source retrieval model. Furthermore, it may serve as a practical alternative to LoRA fine-tuning, particularly when only a limited amount of development data is available.

\section{RELATED WORK}
\subsection{Use of LLMs in Ad-hoc Retrieval}
The application of LLMs has markedly improved ad-hoc retrieval effectiveness~\cite{craswell2021trec}. 
Dense retrieval often utilizes LLMs (e.g., BERT-based models~\cite{devlin2018bert, gao2021condenser}) to encode queries and documents into semantic vectors. 
Various architectures have been proposed, including efficient bi-encoders~\cite{karpukhin2020dense}, more accurate but computationally intensive cross-encoders~\cite{nogueira2019multi}.  
This study focuses on the use of bi-encoder models.

In recent years, foundation models such as Llama2~\cite{touvron2023llama} and Mistral~\cite{jiang2023mistral}, each with 7 billion parameters, have been utilized for ad-hoc retrieval tasks. 
Ma \textit{et al.}~\cite{ma2024fine} constructed RepLLaMA and RankLLaMA by applying LoRA fine-tuning to adapt Llama2 for ad-hoc retrieval tasks. 
Similarly, Wang \textit{et al.}~\cite{wang-etal-2024-improving-text} recently proposed the model e5-mistral-7b-instruct, which was trained with Mistral on synthetic data across various tasks to improve text embeddings. These embeddings have demonstrated strong retrieval effectiveness across a variety of test collections.
We use e5-mistral-7b-instruct as the source retrieval model for model merging.

\subsection{Model Merging} 
Model merging can be categorized into weight-level merging and layer-level merging~\cite{akiba2024evolutionary}. Weight-level merging calculates the parameters of the merged model by combining the parameters of the source models. 
In contrast, layer-level merging constructs new models by selecting layers from the source models, enabling the combination of models with different architectures.

A representative weight-level merging technique involves linearly interpolating the parameters of the source models to compute the parameters of the merged model. Wortsman \textit{et al.}~\cite{wortsman2022model} demonstrated improved performance in image classification tasks by merging multiple models using this method. Other merging techniques, such as SLERP~\cite{shoemake1985animating}, TIES~\cite{yadav2024ties}, DARE~\cite{yu2024language}, and Task Arithmetic~\cite{ilharco2022editing}, have also been proposed. 
In this study, we evaluate the retrieval effectiveness of the merged model using weight-level merging through linear interpolation, which is a simple and widely applicable method. 

\section{METHODOLOGY}
We propose the construction of a new retrieval model via model merging using linear interpolation.

We used the following two models as the source models for model merging. 
The first is a model fine-tuned for retrieval tasks,
which we refer to as the \textit{source retrieval model}.
The second is a model pre-trained on domain-specific documents but lacking retrieval capabilities,
which we refer to as the \textit{source domain-specific model}.
In this study, we used e5-mistral-7b-instruct~\cite{wang-etal-2024-improving-text}, a model fine-tuned from Mistral-7B-v0.1~\cite{jiang2023mistral}, as the source retrieval model. The source domain-specific models will be described in Section~\ref{sec:domains}.

All source models are based on Mistral, which consists of 32 layers. 
It is suggested that different layers in LLMs play distinct roles in language understanding~\cite{meng2022locating}.
To discover a merging strategy that enhances effectiveness in ad-hoc retrieval, we adjusted the merging approach between the two source models depending on the layer. 
Specifically, we divided the model into two segments: the 1st to 16th layers and the 17th to 32nd layers. 
Each segment was assigned a separate hyperparameter to control the weighting of the source models.
The merging was performed using Equations~(\ref{eq:weight_equation1}) and~(\ref{eq:weight_equation2}) for the first and second segments, respectively, as follows:  
\begin{align}
\theta_{\text{lower}}^{\text{(new)}} &= \alpha_{\text{lower}} \theta_{\text{lower}}^{(1)} + \left(1 - \alpha_{\text{lower}}\right) \theta_{\text{lower}}^{(2)} \label{eq:weight_equation1} \\
\theta_{\text{upper}}^{\text{(new)}} &= \alpha_{\text{upper}} \theta_{\text{upper}}^{(1)} + \left(1 - \alpha_{\text{upper}}\right) \theta_{\text{upper}}^{(2)} \label{eq:weight_equation2}
\end{align}
Here, $\theta_{\cdot}^{\text{(new)}}$ denotes the weights of the merged model, $\theta_{\cdot}^{(1)}$ denotes the weights of the source retrieval model, and $\theta_{\cdot}^{(2)}$ denotes the weights of the source domain-specific model.
$\alpha_{\text{lower}}$ and $\alpha_{\text{upper}}$ are hyperparameters.
Once these values are determined, all merged weights $\theta_{\cdot}^{\text{(new)}}$ are directly computed using Equations~(\ref{eq:weight_equation1}) and~(\ref{eq:weight_equation2}).
This process does not involve any backpropagation or iterative updates to the model weights themselves.
Thus, the only values optimized in our method are these two hyperparameters. 
In this study, we employed a grid search to identify the optimal combination of $\alpha_{\text{lower}}$ and $\alpha_{\text{upper}}$. 
We used MergeKit~\cite{goddard2024arcee} for model merging.
The tokenizer and the weights of the token embedding layer were adopted from the source retrieval model.

\section{EXPERIMENTAL SETUP}
\subsection{Overview of Experiments}

\textbf{Experiment 1: Merged Model vs. Source Retrieval Model:} 
This experiment compares the merged model with the source retrieval model, investigating whether model merging can yield a more effective model.\\
\textbf{Experiment 2: Merged Model vs. LoRA Model:}
Fine-tuning is a commonly used approach for building domain-specific retrieval models. In this experiment, we compare the effectiveness of a merged model with that of a fine-tuned source retrieval model. 
When constructing the fine-tuned models, we adopted LoRA (Low-Rank Adaptation). This decision is motivated by the prior research~\cite{ma2024fine}, which suggests that full fine-tuning may be prone to overfitting, as the LoRA model achieves comparable or even superior effectiveness compared to the full fine-tuned model.
Experiment 2 was conducted under two data settings: one with sufficient development data and another with limited development data. The motivation for including the limited data setting is that our proposed method requires learning only two parameters, which is significantly fewer than those in LoRA. As a result, we hypothesize that our method is less prone to overfitting and can outperform LoRA in scenarios with limited development data.

\subsection{Domains and Datasets}
\label{sec:domains}
Each experiment focuses on two domains: the medical domain and a Japanese-language domain.
The medical domain requires specialized knowledge, while we selected Japanese as the language-specific domain due to its structural differences from English and its status as a non-alphabetic language.

\subsubsection{Source Domain-Specific Model}
We used BioMistral-7B~\cite{labrak2024biomistral} as the source domain-specific model for the medical domain.
This model was developed by further pre-training Mistral-7B-Instruct-v0.1 on biomedical corpora.
For the Japanese domain, we used japanese-stablelm-base-gamma-7b~\cite{japanese} as the source domain-specific model.
This model is based on Mistral-7B-v0.1 and has been further pre-trained for Japanese tasks using a corpus of approximately 100 billion tokens. 
It should be noted that these domain-specific models are not trained to output embeddings that are useful for retrieval or general semantic similarity evaluation.

\subsubsection{Datasets}
We utilized two datasets in the medical domain.
The first is NFCorpus~\cite{boteva2016full}, a dataset designed to evaluate the ability to accurately retrieve useful information from biomedical documents. The queries are collected from NutritionFacts.org, and the document corpus consists of biomedical literature collected from PubMed.
The second is SciFact~\cite{wadden-etal-2020-fact}, a dataset designed to evaluate the ability to retrieve documents that support or refute scientific claims. The queries are expert-written scientific claims, and the document corpus consists of abstracts from papers in the fields of science and medicine.

We utilized two datasets in the Japanese domain.
The first is MIRACL~\cite{zhang2023miracl}, which is a dataset designed for multilingual information retrieval, supporting 18 languages. 
We used its Japanese subset, where queries are diverse questions and the document corpus is based on Wikipedia.
The second is JQaRA~\cite{yuichi-tateno-2024-jqara}. 
The task of this dataset is retrieving relevant documents from candidate documents for a given query. 
The queries are question data from JAQKET~\cite{Kurihara_nlp2020}, which is a Japanese QA dataset based on quiz questions. 
The document corpus is based on Wikipedia. 
\subsubsection{Preparation of development and test data}
\begin{table}[t]
\begin{center}
\caption{Statistics of the development and test data. $\dagger$ The document corpus for JQaRA is query-specific.}
\scalebox{0.8}{
   \begin{tabular}{llccc} 
   \toprule
     Domain & Dataset & Split & \#Queries & \#Documents \\
     \midrule
     \multirow{2}{*}{Medical} & \multirow{2}{*}{NFCorpus} & Dev  & $1,000$ & $3,633$ \\
                              &                            & Test & $323$   & $3,633$ \\
     \midrule
     \multirow{2}{*}{Medical} & \multirow{2}{*}{SciFact}  & Dev  & $809$   & $4,725$ \\
                              &                            & Test & $300$   & $5,183$ \\
     \midrule
     \multirow{2}{*}{Japanese} & \multirow{2}{*}{MIRACL}  & Dev  & $1,000$ & $30,321$ \\
                               &                          & Test & $860$   & $8,066$ \\
     \midrule
     \multirow{2}{*}{Japanese} & \multirow{2}{*}{JQaRA}   & Dev  & $1,000$ & $50^\dagger$ \\
                               &                          & Test & $1,667$ & $100^\dagger$ \\
     \bottomrule
   \end{tabular}
   }
 \label{tbl:data_stats}
\end{center}
\end{table}

In this study, we prepared development data and test data for the experiments.

The development data was used to construct the merged models in Experiments 1 and 2, as well as the LoRA model in Experiment 2.
Specifically, it was used for tuning the model merging hyperparameters ($\alpha_{\text{lower}}$ and $\alpha_{\text{upper}}$), and for selecting the number of training epochs and training the models during LoRA fine-tuning (details are provided in Section~\ref{sec:model_construction}).

To construct the development data, we randomly sampled 1,000 queries from the publicly available train split of each dataset. 
Since JQaRA does not provide a train split, we used its dev split instead. 
For SciFact, as the original train set contains fewer than 1,000 queries, we used all 809 available queries.
For each sampled query, the development data included its relevant document(s).
Additionally, for the purpose of creating training data for LoRA, we identified hard negatives by retrieving the top 30 non-relevant documents for each query using BM25. For JQaRA, which already provides hard negatives, we used these provided ones directly.

The test data was used to evaluate the effectiveness of the models constructed in Experiments 1 and 2. In principle, we used the publicly available test split for each dataset. 
Since MIRACL does not provide relevance labels in the test split, we used its dev split as a substitute. 
Furthermore, to reduce the computational cost of evaluation in MIRACL, we limited the target corpus to documents with annotated relevance labels.
While this modification was made for computational efficiency, it is important to note that our results are therefore not directly comparable to evaluations conducted using the full document corpus of MIRACL due to this setup.

Table~\ref{tbl:data_stats} shows the statistics of the development and test data.
Note that we used the same development data for both model merging and LoRA. This is intended to ensure a fair comparison between the two approaches.

\subsubsection{Experimental Conditions for Experiment 2}
As described in Section 4.1, Experiment 2 was conducted under two different data availability conditions: one where sufficient data is available, and the other where only limited data is available. 
The former is referred to as the \textit{full data} setting, in which all development data was used for model merging and LoRA model construction. 
The latter is the \textit{limited data} setting, where a development set of 50 queries, sampled from the full development data, was used for both model merging and LoRA model construction.
To account for sampling variability, we repeated the procedure of sampling 50 queries and evaluating the resulting models 10 times.

\subsection{Model Construction Procedures}
\label{sec:model_construction}
\subsubsection{Model Merging}
To explore the optimal balance between the two source models, we performed a grid search over the hyperparameters $\alpha_{\text{lower}}$ and $\alpha_{\text{upper}}$.
Each hyperparameter could take values of  0.00, 0.25, 0.50, 0.75, and 1.00, leading to a total of 25 possible configurations.
However, two of these configurations represent configurations equivalent to the original unmerged models: when both $\alpha_{\text{lower}}$ and $\alpha_{\text{upper}}$ are 1.00, the model is equivalent to the source retrieval model (e5-mistral-7b-instruct), and when both $\alpha_{\text{lower}}$ and $\alpha_{\text{upper}}$ are 0.00, it corresponds to the source domain-specific model (BioMistral-7B or japanese-stablelm-base-gamma-7b).
These two configurations were therefore excluded from our grid search.
As a result, we evaluated the remaining 23 unique configurations.
For each of the 23 configurations, we computed the nDCG@10 score on the development set. The pair of $\alpha_{\text{lower}}$ and $\alpha_{\text{upper}}$ that yielded the highest nDCG@10 was selected as the optimal setting for model merging.

\subsubsection{LoRA Fine-tuning}
For LoRA fine-tuning, the number of training epochs was treated as a key hyperparameter to be optimized. 
Other hyperparameters, such as learning rate, were set based on prior works~\cite{ma2024fine,wang-etal-2024-improving-text}. 
To determine the optimal number of epochs, we first split the development data into a training subset (dev-training) and a validation subset (dev-validation) using an 80:20 ratio. 
We then trained models with early stopping, monitoring performance on the dev-validation set to identify the best epoch. 
Finally, the LoRA model was retrained using the entire development data for this optimal number of epochs to produce the final model. 
All LoRA fine-tuning was conducted with Tevatron~\cite{Gao2022TevatronAE}.

\section{EXPERIMENTAL RESULTS}
\begin{table}[t]
\centering
\caption{
nDCG@10 for source retrieval and merged models. $*$ indicates significant difference. Hyperparameters selected on development data are also shown.}
\label{tab:merged}
\scalebox{0.95}{
\begin{tabular}{lccccc}
\toprule
\textbf{Dataset} & \textbf{Source} & \textbf{Merged}  & \(\alpha_{\text{lower}}\) & \(\alpha_{\text{upper}}\)\\
\hline
NFCorpus    & 39.02 & \textbf{40.59*} & 0.75 & 1.00\\
SciFact     & 77.32 & \textbf{77.63}  & 0.75 & 1.00\\
\midrule
MIRACL      & 75.49 & \textbf{77.90*} & 0.75 & 0.75\\
JQaRA       & 60.80 & \textbf{63.89*} & 0.50 & 1.00\\
\bottomrule
\end{tabular}
}
\end{table}
For both experiments, we assessed statistical significance using a paired t-test with a significance level of 5\%.

\paragraph{Experiment 1: Merged Model vs. Source Retrieval Model}
Table~\ref{tab:merged} shows the nDCG@10 scores of the source retrieval and merged models for each dataset. 
It also includes the hyperparameter values used for model merging, which were selected based on the development data of each dataset.
From Table~\ref{tab:merged},  we observe that the merged models significantly outperformed the source retrieval model on three out of four datasets.
These results suggest that model merging can be a viable approach to constructing domain-specific retrieval models that outperform the original source retrieval model.

Analyzing the selected hyperparameters further, we observe that the optimal values of $\alpha_{\text{lower}}$ and $\alpha_{\text{upper}}$ were predominantly found in the higher range (e.g., 0.75–1.0), indicating that assigning greater weight to the retrieval model tends to yield superior effectiveness.

\paragraph{Experiment 2: Merged Model vs. LoRA Model}
Table~\ref{tab:merged_results} presents the nDCG@10 scores of LoRA models and merged models across four datasets under both full and limited data settings. 
In the limited data setting, scores are averaged over 10 runs, with standard deviations shown in parentheses.

\begin{table}[t]
\centering
\caption{nDCG@10 for LoRA and merged models in full and limited data settings. $*$ indicates significant difference.}
\label{tab:merged_results}
\scalebox{0.78}{
\begin{tabular}{lcccccc}
\toprule
 & \multicolumn{3}{c}{\textbf{Full Data}} & \multicolumn{3}{c}{\textbf{Limited Data (average of 10 runs)}} \\
\cmidrule(lr){2-4} \cmidrule(lr){5-7}
\textbf{Dataset} & \textbf{LoRA} & \textbf{Merged} & \textbf{Diff} & \textbf{LoRA} & \textbf{Merged} & \textbf{Diff} \\
\midrule
NFCorpus    & \textbf{41.62} & 40.59 & -1.03 & 36.09(4.56) & \textbf{40.36*(0.72)} & +4.27 \\
SciFact     & \textbf{85.76*} & 77.63 & -8.13 & \textbf{77.51(1.46)} & 77.31(0.76) & -0.20 \\
\midrule
MIRACL      & 74.83 & \textbf{77.90*} & +3.07 & 72.57(2.88) & \textbf{76.60*(1.28)} & +4.03 \\
JQaRA       & 62.86 & \textbf{63.89*} & +1.03 & 59.35(2.85) & \textbf{64.16*(0.59)} & +4.81 \\
\bottomrule
\end{tabular}
}
\end{table}

In the full data setting, the effectiveness difference between LoRA and merged models varied across datasets.
LoRA outperformed merged models on NFCorpus and SciFact, with the difference in SciFact being statistically significant. 
In contrast, the merged models significantly outperformed LoRA on MIRACL and JQaRA. 
These results suggest that when sufficient data is available for model construction, the effectiveness of LoRA and model merging depends on the dataset and the choice of source domain-specific models used for merging.

In the limited data setting, the merged models achieved significantly higher average nDCG@10 scores than LoRA models on NFCorpus, MIRACL, and JQaRA.
This performance advantage over LoRA was more substantial compared to the full data setting; for instance, NFCorpus saw a reversal where the merged model surpassed LoRA, and the performance gaps on MIRACL and JQaRA widened in favor of the merged models.
Moreover, across all datasets, the merged models consistently exhibited smaller standard deviations. 
These observations suggest that while LoRA's effectiveness can exhibit high variability depending on the particular training instances when trained with limited data, model merging offers a more consistently effective and robust solution. 
This highlights model merging's potential under data-scarce conditions. 
A primary reason may be that our approach tunes only two parameters, far fewer than LoRA, making it less prone to overfitting.

\section{CONCLUSION AND FUTURE DIRECTIONS}
We applied model merging to ad-hoc retrieval tasks and demonstrated its effectiveness. 
The experimental results demonstrated that the merged models outperformed the source retrieval model on three out of four datasets.
Furthermore, particularly when limited data is available for model construction, our findings suggest that model merging has the potential to construct more stable models with better retrieval effectiveness compared to LoRA fine-tuning.

This study represents an initial step in exploring the application of model merging to ad-hoc retrieval tasks, and several interesting research directions remain, including:
(1) Exploring more advanced model merging techniques beyond simple linear interpolation. This includes various existing methods~\cite{shoemake1985animating,yadav2024ties,yu2024language,ilharco2022editing} and approaches like \textit{evolutionary model merging}~\cite{akiba2024evolutionary}, which optimizes hyperparameters based on effectiveness metrics of the specified task. 
Such explorations could uncover optimal merging strategies and yield greater improvements in retrieval effectiveness.
(2) Understanding the interplay between source models and target corpora is key to optimizing model merging. The inconsistent effectiveness across datasets, exemplified by the limited improvement on SciFact, highlights the need to identify what specific model and corpus characteristics lead to successful merging. This knowledge would inform guidelines for more effective merging strategies.

\begin{acks}
This work was supported by JSPS KAKENHI Grant Numbers JP24K03228, JP25K03228 and JP25K03229, and by ROIS NII Open Collaborative Research 2025-251S4-22794. We thank Tokyo ACM SIGIR Chapter for the discussion at POWIR.
\end{acks}

\section*{GenAI Usage Disclosure}
We used OpenAI ChatGPT and Google Gemini only for editing and polishing the author-written text. No new content was generated by these tools. All content remains the authors' responsibility. 

\bibliographystyle{ACM-Reference-Format}
\balance
\bibliography{reference}

\end{document}